\documentclass[12pt,draftclsnofoot,onecolumn]{IEEEtran}

\usepackage{amsfonts}
\usepackage{times}
\usepackage{latexsym}
\usepackage{amssymb}
\usepackage{amsmath}
\usepackage{cite}
\usepackage{verbatim}

\newcommand{\bydef}{\triangleq}

\def\SNR{{\textsf{SNR}}}

\def\bydef{:=}

\def\bb0{{\mathbb{0}}}

\def\bydef{:=}

\def\bb{{\mathbf{b}}}

\def\bh{{\mathbf{h}}}

\def\b0{{\mathbf{0}}}

\def\bbC{{\mathbb{C}}}

\def\bbE{{\mathbb{E}}}

\def\bbN{{\mathbb{N}}}

\def\bbR{{\mathbb{R}}}

\def\bydef{:=}

\def\sf0{{\mathsf{0}}}

\pdfoutput=1
\usepackage{graphicx}
\usepackage{amssymb}
\usepackage{amsfonts}
\usepackage{amsmath}
\usepackage{latexsym}

\begin{document}
\newtheorem{thm}{Theorem}
\newtheorem{lemma}{Lemma}
\newtheorem{rem}{Remark}
\newtheorem{exm}{Example}
\newtheorem{prop}{Proposition}
\newtheorem{defn}{Definition}
\def\proof{\noindent\hspace{0em}{\itshape Proof: }}
\def\endproof{\hspace*{\fill}~\QED\par\endtrivlist\unskip}
\def\bh{{\mathbf{h}}}
\def\SNR{{\mathsf{SNR}}}
\title{Throughput-Delay-Reliability Tradeoff with ARQ in Wireless Ad Hoc Networks
}
\author{Rahul~Vaze \\
Tata Institute of Fundamental Research\\
School of Technology and Computer Science\\
Homi Bhabha Road, Mumbai 400005\\
Email: vaze@tcs.tifr.res.in}

\date{}
\maketitle
\noindent
\begin{abstract}
Delay-reliability (D-R), and  throughput-delay-reliability (T-D-R) tradeoffs in an ad hoc network are derived for single hop and multi-hop transmission with automatic repeat request (ARQ) on each hop. The delay constraint is modeled by assuming that each packet is allowed at most $D$ retransmissions end-to-end, and the reliability  is defined as the probability that the packet is successfully decoded in at most $D$ retransmissions. The throughput of the ad hoc network is characterized by the transmission capacity, which is defined to be the maximum allowable density of transmitting nodes satisfying a per transmitter
receiver rate, and an outage probability constraint, multiplied with the rate of transmission and the success probability. Given an end-to-end retransmission constraint of $D$, the optimal allocation of the number of retransmissions allowed at each hop  is derived that maximizes a lower bound on the transmission capacity. Optimizing over the number of hops, single hop transmission is shown to be optimal for maximizing a lower bound on the transmission capacity  in the sparse network regime. 
\end{abstract}

\section{Introduction}
The transmission capacity of an ad hoc network is  the maximum allowable
 density of transmitting nodes, satisfying a per transmitter receiver rate,
and outage probability constraints \cite{Weber2005,Weber2007, Weber2008, Baccelli2006}. 
The transmission capacity is computed under the assumption that the transmitter locations are distributed
as a Poisson point process (PPP) using tools from stochastic geometry \cite{Weber2005,Weber2007, Weber2008, Baccelli2006}. 
The transmission capacity framework allows for tractable analysis with different
physical layer transmission techniques, such as use of multiple antennas
\cite{Hunter2008, Huang2008, Vaze2009l, Jindal2008a}, bandwidth partitioning \cite{Jindal2007},
and successive interference cancelation \cite{Weber2007}.

Most of the prior work on computing the transmission capacity of ad hoc networks has been limited to single hop communication. Recently,  under some assumptions, \cite{Andrews2009} computed the transmission capacity of ad hoc network with multi-hop transmissions, and automatic repeat request (ARQ) on each hop. To account for retransmissions and multiple hops, \cite{Andrews2009} normalized the transmission capacity by end-to-end expected delay, and defined the success event as the event that  the packet is successfully decoded in at most $D$ retransmissions. Modeling $D$ as delay, the relationship between the success probability and $D$ captures the delay-reliability (D-R) tradeoff, while the transmission capacity expression characterizes the throughput-delay-reliability (T-D-R) tradeoff. Some other related papers on multi-hop networks include \cite{Stamatiou2009, Haenggi2005b, Sikora2006}, where \cite{Stamatiou2009} computes the optimal number of hops that minimize the end-to-end delay, while \cite{Haenggi2005, Sikora2006} derive the optimal number of hops in a line network with no interference.

The analysis  carried out in \cite{Andrews2009} assumes $D \rightarrow \infty$, and independent packet success/failure events across time slots. The  second assumption can only be justified for very low density of transmitters, and does not hold true otherwise \cite{Ganti2009}. 
The result of \cite{Andrews2009} is useful in determining the optimal number of hops that maximize an upper bound on the transmission capacity with no retransmissions constraint, however, does not characterize the D-R or the T-D-R tradeoff of the ad hoc network.


To characterize the D-R and T-D-R tradeoffs, in this paper, we  derive an exact expression for the transmission capacity with multiple hops and retransmissions. In contrast to \cite{Andrews2009} to derive the transmission capacity, we 
i)  use a finite $D$, 
ii) do not assume independence of success/failures of packets across time slots. 
iii)  assume that each transmitter retransmits using the slotted ALOHA protocol. Our results are summarized as follows.
\begin{itemize}
\item We derive the exact expressions for the success probability, and the transmission capacity for single hop and multi-hop transmissions with finite $D$. 
\item The exact expressions are quite complicated, and to obtain more insights we derive tight upper and lower bounds on the success probability using the FKG inequality \cite{Grimmett1980}.
\item Using the derived bounds, we characterize the D-R, and the T-D-R tradeoff in an ad hoc network. We show that the success probability increases as $1-x^{D+1}$ ($x<1$ is a constant) for single hop transmission. 
\item For equidistant hops we show that equally distributing the total retransmission constraint among all the hops is optimal for  maximizing a lower bound on the transmission capacity.
\item For multiple equidistant hops, we  derive the optimal number of hops that maximize a lower bound on the  transmission capacity. In the sparse network regime we show that it is optimal to transmit over a single hop. 

\end{itemize}

\section{System Model}
\label{sec:sys}
Consider an ad hoc network where multiple source destinations pairs want to communicate with each other without any centralized control. The location of each source node ${\cal S}_m, \ m\in \bbN$ is modeled as a homogenous Poisson point
process (PPP) on a two-dimensional plane with intensity $\lambda_0$ \cite{Stoyan1995} \footnote{Our model precludes mobility of nodes, and is restricted to averaging with respect to the PPP spatial node distribution.}. We assume that source ${\cal S}_m$ is located at a distance $d$ from its intended receiver ${\cal D}_m$ $\forall m$, with $N-1$ relays $R_{nm}$, $n=1,2,\ldots,N-1$ ($N$ hops) in between\footnote{The results of this paper can be generalized for random distances between the source and the destination.}. The $n^{th}$ hop distance is $d_{n}$, such that $\sum_{n=1}^Nd_{n} = d$. Thus  link $m$ is described by the set of nodes
$\{{\cal S}_m, {\cal R}_{1m}, \ldots, {\cal R}_{N-1m}, {\cal D}_m\}$. We assume that all nodes in the network have single antenna each. The transmission happens hop by hop using the decode and forward strategy. 
We consider ARQ on each hop, where the receiver informs the transmitter of the success (ack) or failure (nack) of the packet decoding instantly, and without any errors. 
We assume that at most $D$ end-to-end retransmissions are allowed between ${\cal S}_m$ and ${\cal D}_m, \ \forall \ m$. This requirement is used to model the delay constraint, which gives rise to the outage event that  the packet is not successfully decoded at the destination after $D$ retransmissions. Let $D_n$ be the number of retransmissions used on hop $n$, then $D = \sum_{n=1}^ND_n$. For simplicity, same packet is assumed to be retransmitted (at most $D$ times) with every nack, without any incremental redundancy or rate adaptation.

Following \cite{Andrews2009}, we assume that there is only one active packet on each link\footnote{ For more discussion on this assumption see Remark 2 \cite{Andrews2009}.}, i.e. the source waits to transmit the next packet until the previous packet has been  received by the destination, or the delay constraint has been violated.  Let the transmitter and receiver on link $m$ in time slot $t$ be $T_m^t$ and $R_m^t$, respectively, $T_m^t \in \{{\cal S}_m, {\cal R}_{1m}, \ldots, {\cal R}_{N-1m}\}$, $R_m^t \in \{{\cal R}_{1m}, \ldots, {\cal R}_{N-1m},{\cal D}_m\}$. Then the set of interfering nodes for $R_m^t$ is $\Phi_t^m \bydef \{\Phi_t \backslash T_m^t\}$, where $\Phi_t \bydef \{T_k^t, \ k\in \bbN\}$. Using the Slivnyak's Theorem, the stationarity of the PPP, and the random translation invariance property of the PPP \cite{Stoyan1995, Daley2003}, the locations of interferers of $\Phi_t^m$ are distributed as a PPP with intensity $\lambda_0, \ \forall \ t, m$ \cite{Andrews2009}.

We consider a slotted ALOHA like random access protocol, where each transmitter (source or any relay) attempts to transmit its packet with an access probability $p$, independently of all other transmitters. Consequently, the active transmitter process is also a homogenous
PPP on a two-dimensional plane with intensity
$\lambda \bydef p\lambda_0$.
Note that when the active transmitter process is a PPP, the success/failure of packet decoding at different receivers is correlated \cite{Ganti2009}. 
Therefore retransmission of packets depending on the nack introduces correlation among the active transmitters process, and it is no longer a random thinning of PPP, and consequently not a PPP. Violating the PPP assumption, however, entails  analytical intractability. 
To satisfy the PPP assumption on the active transmitter locations, we assume that similar to the newly arrived packets in its queue, each transmitter uses a slotted ALOHA protocol with access probability $p$ to retransmit old packets as well.

For the purpose of analysis we consider a typical link $\{{\cal S}_0, {\cal R}_{10}, \ldots, {\cal R}_{N-10}, {\cal D}_0\}$. It has been shown in \cite{Weber2005} that for the PPP distributed transmitter locations, the performance of the typical source destination pair
is identical to the network wide performance.
For simplicity we refer to link $\{{\cal S}_0, {\cal R}_{10}, \ldots, {\cal R}_{N-10}, {\cal D}_0\}$ as $\{{\cal S}_0, {\cal D}_0\}$.
Let  $n^{th}$  relay ($n=0$ corresponds to the source ${\cal S}_0$) be the active transmitter for the typical link $\{{\cal S}_0, {\cal D}_0\}$ at time slot $t$, i.e. $T_0^t = R_{n0}$. Then the received signal at the $n+1^{th}$ relay (defined $R_0^t$) of link $\{{\cal S}_0, {\cal D}_0\}$ at time slot $t$ is
\begin{eqnarray}\label{eq:rxsig}
y^t_0 = \sqrt{P}d^{-\alpha/2}h^{nt}_{00}x^t_0 + \sum_{T_s^t \in \Phi_t^m }\sqrt{P}{\mathbf 1}_{T_s^t}d_s^{-\alpha/2}h^{nt}_{0s}x^t_s + z^t_0,
\end{eqnarray}
where $P$ is the transmit power of each transmitter,
$h^{nt}_{0s}\in \bbC$
is the channel coefficient between $T_s^t$ and $R_0^t$ on hop $n$, $d_s$ is the distance between $T_s^t$ and $R_0^t$, $\alpha$ is the path loss exponent $\alpha > 2$,
$ x^t_s \sim  {\cal CN}(0,1)$ is the signal transmitted from $T_s^t$ in time slot $t$,  ${\mathbf 1}_{T_s^t}=1$ with probability $p$, and $0$ otherwise, due to ALOHA transmission strategy, and $z^t_0$ is the additive white Gaussian noise. All
results in this paper are valid for $\alpha >2$.
We consider the interference limited regime, i.e.
noise power is negligible compared to the interference power, and
henceforth drop the noise contribution \cite{Weber2005}\footnote{This assumption is made for simplicity of exposition, and all results of this paper can be easily extended to the case of  additive noise as well. Footnote $10$ on Page $14$  explicitly describes how to extend results of Section \ref{sec:optN} for additive noise case.}.
We also assume $P=1$, since the signal to interference ratio (SIR) is independent of $P$.
We assume that each $h^{nt}_{0s}$ is
independent and identically distributed complex normal random variable with mean zero and variance $1$ $({\cal CN}(0,1))$  $\ \forall \ n, t, s$.

Let $SIR^n_t$ denote the $SIR$ on hop $n$ of link $\{{\cal S}_0, {\cal D}_0\}$ at time slot $t$. 
With the received signal model (\ref{eq:rxsig}),
$SIR_{t}^n \bydef \frac{ 
d_n^{-\alpha} |h_{00}^{nt}|^2
}
{
\sum_{
T_s \in \Phi^t_{n} \backslash \{T_0\}
}
{\mathbf 1}_{T_s} 
d_{T_s}^{-\alpha}|h_{0s}^{nt}|^2}$. We assume that the rate of transmission for each hop is $R(\beta) = \log(1+\beta)$ bits/sec/Hz, therefore,  a packet transmitted by $T_0^t$ can be successfully decoded at $R_0^t$ in  time slot $t$ on hop $n$, if $SIR_{t}^n \ge \beta$.
%
Let $M_n$ be  the random variable denoting the number of transmissions used at hop $n$,  $M_n\le D_n+1$. Then the expected delay on the $n^{th}$ hop is $\bbE\{M_n\} $. Let $P_s$ be the probability that the packet is successfully decoded by the destination ${\cal D}_0$ within $D$ retransmissions.  Then the transmission capacity of ad hoc network with multi-hop transmission is defined as\begin{equation}
\label{tcmulti}
C \bydef \max_{D_n, \ \sum_{n=1}^N{D_n}\le D}\frac{P_s \lambda R}{\bbE\{\sum_{n=1}^NM_n\}} \ \text{bits/sec/Hz/m$^2$},
\end{equation}
where in contrast to \cite{Andrews2009} we have not multiplied the transmission distance $d$. 
The transmission capacity quantifies the end-to-end rate that can be supported by $\lambda$ simultaneous transmissions/unit area, with outage probability $P_s$, and maximum delay $D+N$. Thus, the transmission capacity captures the {\bf T-D-R tradeoff} of ad hoc networks, where  throughput $=C$, maximum delay $\le D+N$, and reliability $=P_s$. Similarly, the definition of $P_s$ captures the {\bf D-R tradeoff} of ad hoc network. 
\section{Single Hop Transmission}
\label{sec:singlehop}

In this section we consider  a single hop ad hoc network $N=1$. Our goal in this section is to derive $P_s$ and $C$, when at most $D$ retransmissions are allowed for each packet. Towards that end, 
let $P_s^j$ be the  probability of success in the $j^{th}$ time slot. Then 
\[P_s^j = P(\cup_{k=0, 1, 2, \ldots,   j-1} \{\text{failures in any $k$ time slots, \ success in the $j^{th}$ time slot}\}),\] since at each time slot, retransmission happens only with probability $p$\footnote{If $p=1$,  i.e. retransmission happens  with each nack, then $P_s^j = P(\{\text{failures in  $j-1$ time slots, \ success in the $j^{th}$ time slot}\})$. }. Clearly, the events $\cup_{k=0, 1, 2, \ldots,   j-1} \{\text{failures in any $k$ time slots, \ success in the $j^{th}$ time slot}\}$ are mutually exclusive  for any $j$, $j=1,\ldots,D+1$, hence, the success probability $P_s$ is  
\begin{eqnarray}\label{psucdef1}
P_{s} &= &   \sum_{j=1}^{D+1} P_s^j.
\end{eqnarray} 
Note that $P_s \rightarrow 1$ as $D\rightarrow \infty$. In this section we only consider $N=1$, and hence  drop the hop index $n$ from all parameters, e.g. $SIR_t^n$ is denoted as $SIR_t$. Since $SIR_t$ is identically distributed $\forall \ t$, $P\left(\text{success in the $j^{th}$ time slot}\right)$ only depends on how many failures have happened before time slot $j$, and not where those failures happened\footnote{For example, $P\left(SIR_1 < \beta, SIR_{t}< \beta\right) = P\left(SIR_1 < \beta, SIR_{n}< \beta\right)$, $t\ne n$, since the channel coefficients are independent across time slots, and in any time slot each transmitter transmits with probability $p$ independent of others.}. 
Therefore it follows that
\begin{eqnarray} \label{psucdef2}
P_s^j& =& \sum_{k=0}^{j-1}  \ {j-1 \choose k} p^k(1-p)^{j-1-k} p  P\left(
SIR_1 <  \beta, 
\ \ldots,  \ SIR_{k} <  \beta,
 \  SIR_j \ge \beta  \right),\end{eqnarray}
by accounting for $k=0$, or $1$, or , $\ldots, j-1$ failures before success at the the $j^{th}$ slot. 
Computing the joint probability in (\ref{psucdef2}),  $P_s$ is given by the following Proposition.
 \begin{prop}
 \label{psucsingle} The success probability $P_s$ is given by 
   \begin{eqnarray*}
 P_s &=&  \sum_{j=1}^{D+1} \sum_{k=0}^{j-1}  \ {j-1 \choose k} p^k(1-p)^{j-1-k} p\sum_{\ell=0}^{k} (-1)^{\ell}{k \choose \ell }  \times \\ 
 &&\exp\left( -\lambda \int_{\bbR^2} 1- \left(\frac{p}{1+ d^{\alpha}\beta x^{-\alpha}} + 1-p \right)^{\ell+1} {\mathbf d}x \right).
 \end{eqnarray*}
 \end{prop}
  \begin{proof} See Appendix \ref{app:jointprob}.
\end{proof}

Recall that $M$ is the random variable denoting the number of retransmissions required. Note that $M$ takes values in $[0:D+1]$ with probability 
$P(M=j) = P_s^j, \ j=0,1,2,\ldots, D$, and $P(M=D+1) = P_s^{D+1} + \sum_{j=0}^{D}P_s^j = P_s^{D+1} +(1-P_s)$. The second term in $P(M=D+1)$ is to account for delay incurred by packets that are not decoded even after $D$ retransmissions.
Using the derived expression for $P_s^j$ in Appendix \ref{app:jointprob}, the expected number of retransmissions $\bbE\{M\}$ is computed as follows.
 \begin{prop}\label{prop:expdelsingle} The expected delay $\bbE\{M\}$ in a single hop ad hoc network with at most $D$ retransmissions is
   \begin{eqnarray*}
 \bbE\{M\} &=&  \sum_{j=1}^{D+1} \sum_{k=0}^{j-1} \sum_{\ell=0}^{k}  j {j-1 \choose k} p^{k+1}(1-p)^{j-1-k} (-1)^{\ell}{k \choose \ell }  \times \\ 
 &&\exp\left( -\lambda \int_{\bbR^2} 1- \left(\frac{p}{1+ d^{\alpha}\beta x^{-\alpha}} + 1-p \right)^{\ell+1} {\mathbf d}x \right) + (D+1)(1-P_s).
 \end{eqnarray*}
 \end{prop}
 \begin{thm}\label{tcsingle} The transmission capacity of a single hop ad hoc network with at most $D$ retransmissions is 
 $C = \lambda R \frac{P_s}{\bbE\{M\}}$,
     where $P_s$ is given by Proposition \ref{psucsingle}, and $\bbE\{M\}$ is given by Proposition \ref{prop:expdelsingle}. 
 \end{thm}
 
Proposition \ref{psucsingle} and Theorem \ref{tcsingle} give an exact expression for the success probability, and the transmission 
capacity, respectively, of an ad hoc network with single hop transmission, and retransmissions constraint of $D$. 
Because of the correlation of $SIR's$ across different time slots with PPP distributed transmitter locations, the derived expressions are complicated, and do not allow for a simple closed form expression for $P_s$, and $C$, as a function of $D$.
To get more insights on the dependence of $P_s$, and $C$, on $D$ (to obtain simple D-R and T-D-R tradeoffs), we next derive tight lower and upper bounds on $P_s^j$, and consequently on $P_s$, and the transmission capacity $C$.
\subsection{Bounds On The Transmission Capacity}
For deriving the bounds we need the following definitions.
\begin{defn}\label{exmdecevent} Let $(\Omega, {\cal F}, {\cal P}$) be the probability space. Let $A$ be an event in ${\cal F}$, and ${\mathbf 1}_{A}$ be the indicator function of $A$. Then the event $A \in \ {\cal F}$ is called increasing if ${\mathbf 1}_{A}(\omega) \le {\mathbf 1}_{A}(\omega')$, whenever $\omega \le \omega'$ for some partial ordering on $\omega$. The event $A$ is called decreasing if its complement $A^{c}$ is increasing.

\end{defn}
 
\begin{exm}\label{exm:incevent} Success event $\{SIR >\beta\}$ is a decreasing event. Follows by considering $\omega = (a_1, a_2, \ldots, )$ where for $n \in \bbN$, $a_n =1$ if $T_m^t$ is active, $0$ otherwise, and the definition of $SIR_t$.

\end{exm}

\begin{lemma}\label{lemfkg}(FKG Inequality \cite{Grimmett1980}) If both $A, B \in {\cal F}$ are increasing or decreasing events then $P(AB) \ge P(A)P(B)$.

\end{lemma}

{\bf Upper bound on the success probability $P_s$.}
\begin{prop}
\label{prop:upboundpsuc}
The success probability $P_s$ with single hop transmission in an ad hoc network with at most $D$ retransmissions  is upper bounded by 
$
P_s  \le 1- (pq+1-p)^{D+1}$,
where $q\bydef P\left(SIR_1<  \beta  \right) = 1-\exp \left(-\frac{\lambda 2 \pi^2 d^2 \beta^{
 \frac{2}{\alpha}}Csc\left(\frac{2\pi}{\alpha}\right)}{\alpha}\right)$ \cite{Baccelli2006}.
 \end{prop}
 \begin{proof} See Appendix \ref{app:upboundPs}.
\end{proof}

{\bf Lower bound on the success probability $P_s$.}
\begin{prop}
The success probability $P_s$ with single hop transmission in an ad hoc network with at most $D$ retransmissions is lower bounded by 
\begin{eqnarray} \label{lbpsuc}
P_s&\ge& P\left(SIR_{D+1} \ge \beta \ | \ SIR_1 <  \beta, 
\ \ldots,  \ SIR_{D} <  \beta \right) 
 \frac{1- (pq+1-p)^{D+1}}{1-q}. 
  \end{eqnarray}
\end{prop}
\begin{proof} See Appendix \ref{app:lboundPs}.
\end{proof}

For small values of $D$ we can analytically show that  
$P\left(SIR_{D+1} \ge \beta \ | \ SIR_1 <  \beta, 
\ \ldots,  \ SIR_{D}\right.$ $\left. <  \beta \right) \approx 1-q$, and hence our derived bounds on $P_s$ are tight. For higher values of $D$ also, the bounds can be shown to be tight using simulations in the sparse network regime i.e. small $\lambda$ or $\lambda \rightarrow 0$. Thus, from here on in this paper we assume that $P_s = c \left(1-(pq+1-p)^{D+1}\right)$, where $c<1$ is a constant.
{\bf D-R tradeoff:} From the upper and lower bound, 
\begin{eqnarray}\label{eq:pucsingle}
P_s &=& c(1- (1-p+pq)^{D+1}).\end{eqnarray}
 Thus the success probability increases as $1-x^{D+1}$ with $D$, where $x<1$ is a constant.
 Using the derived upper and lower bound the expected delay is 
 \begin{eqnarray} \label{expdelay}
\bbE\{M\}&=&  c\left[\frac{1-(pq + 1-p)^{D+1}}{(1-q)}\right] + (D+1)(1-c).
\end{eqnarray}
 {\bf T-D-R tradeoff:}  Using the derived expression for $P_s$ (\ref{eq:pucsingle}), and $\bbE\{M\}$ (\ref{expdelay}), we get \[C = \frac{c(1-(pq + 1-p)^{D+1})\lambda R}{c\left[\frac{1-(pq + 1-p)^{D+1}}{(1-q)}\right] + (D+1)(1-c)} \  \text{bits/sec/Hz/$m^2$}. \] 
%
 
 {\it Discussion:} 
 In this section we derived the exact D-R, and the T-D-R tradeoffs of a single hop ad hoc network. The exact expressions  are  fairly complicated, and do not yield a simple relationship between $C$, $P_s$, and $D$, for any arbitrary $D$. To obtain more meaningful insights on the relationship between $P_s$, $D$, and $C$, we derived tight upper and lower bounds on the success probability, and showed that the bounds are tight for small $D$, or in the sparse network regime. The bounds reveal that even though the success/failure of packet decoding is correlated across time slots, for small $D$ or in a sparse network, the success probability $P_s$ is equal to a $c P_s^{indep}$, where $c<1$ is a constant, and $P_s^{indep}$ is the success probability in less than $D$ retransmissions if the success/failure of packet decoding is independent across time slots.

 \section{Multi-Hop Transmissions}
 \label{sec:multi-hop}
In this section we consider multi-hop communication (arbitrary $N$). 
We analyze the case of $N=2$, $N >2$ follows similarly.
With at most $D_1$ retransmissions on the first hop, and $D_2$ retransmissions on the second hop ($D_1+D_2 =D$), the success probability $P_{s}$ for transmission
between ${\cal S}_0$ and ${\cal D}_0$ is  
$P_{s} =  \sum_{j=1}^{D_1+1}\sum_{k=1}^{D_2+1} P_s^{jk}$,
where $P_s^{jk} = P(\text{success in the $j^{th}$ time slot on hop $1$}$, $\text{ success in the $k^{th}$ time slot on hop $2$})$. 
Let ${\cal E}_{jk}$ be the event $\{\text{success in the $j^{th}$ time slot on hop}$ $\text{$1$, success in the $k^{th}$ time slot on hop $2$}\}$, and 
${\cal F}_{sj}^n$ be the event $\{\text{failures in any $s$ time slots on}$ $\text{ hop $n$, \ success in the $j^{th}$ time slot on hop $n$, $s<j$}\}$. Then ${\cal E}_{jk} =  \cup_{\ell, m} \left\{ {\cal F}_{\ell j}^1 \cap {\cal F}_{mk}^2\right\}$, $\ell =0, 1,$ $  \ldots,   j-1, \ m =0,1,2,\ldots, k-1$, since at each time slot, retransmission happens only with probability $p$ at each hop.  

Note that $SIR^n_j$ is identically distributed $\forall \ j$, thus, $P_s^{jk}$ only depends on how many failures have happened before time slot $j$ on hop $1$, and time slot $k$ on hop $2$, respectively, and not where those failures happened. 
Therefore expanding $P_s^{jk}$,
{\begin{eqnarray}\nonumber 
P_{s} 
 &=&
\sum_{j=1}^{D_1+1} \sum_{k=2}^{D_2+1} \sum_{\ell=0}^{j-1}  \ {j-1 \choose \ell}p^{\ell}(1-p)^{j-1-\ell} \sum_{m=0}^{k-1}  \ {k-1\choose m}p^m(1-p)^{k-1-m} p^2 \times \\ \label{psucdummy} 
&&P\left(
SIR^1_1 <  \beta, \  
\ \ldots  SIR^1_{\ell} <  \beta, 
 \  SIR^1_j \ge \beta,   SIR^2_1 <  \beta, \  
\ \ldots  SIR^2_{m} <  \beta,
 \  SIR^2_{k} \ge \beta\right).  \end{eqnarray}
 Computing the joint probability, $P_s$ is given by the next proposition.

 \begin{prop}
 \label{jointprobmulti}
  \begin{eqnarray*}
  P_s  & = &  \sum_{j=1}^{D_1+1} \sum_{k=2}^{D_2+1} \sum_{\ell=0}^{j-1}  \ {j-1 \choose \ell}p^{\ell}(1-p)^{j-1-\ell} \sum_{m=0}^{k-1}  \ {k-1\choose m}p^m(1-p)^{k-1-m} p^2 \\ 
  &&\times\sum_{r=0}^{\ell} (-1)^{r}{k \choose r } \sum_{s=0}^{m} (-1)^{s}{m \choose s }  \times \\ 
 \ \ \ \ \ \ && 
\exp\left( -\lambda \int_{\bbR^2} 1- \left(\frac{p}{1+ d_1^{\alpha}\beta x^{-\alpha}} + 1-p \right)^{r+1} 
\left(\frac{p}{1+ d_2^{\alpha}\beta x^{-\alpha}} + 1-p \right)^{s+1} {\mathbf d}x \right)\end{eqnarray*}
 \end{prop}
\begin{proof}
 Proof is similar to Proposition \ref{psucsingle}.
\end{proof}
The expected delay $\bbE\{M\}$ for $N=2$ can be computed easily  by using the linearity of expectation, since 
$\bbE\{M\} = \bbE\left\{M_1\right\} + \bbE\left\{M_2\right\}$, where $\bbE\left\{M_n\right\}$ is given by Proposition \ref{prop:expdelsingle}.

\begin{thm}
\label{thm:tcmultiexact} The transmission capacity of an  ad hoc network with $N=2$-hop transmission and end-to-end retransmission constraint of $D$ is 
$C = \max_{D_n, \ \sum_{n=1}^N D_n \le D}\frac{\lambda R P_s}{\sum_{n=1}^2\bbE\left\{M_n\right\}} \ \ \ \ \text{bits/sec/Hz/$m^2$}$,
where $P_s$ is given by (\ref{psucdummy}) and Proposition \ref{jointprobmulti}, and  $\bbE\left\{M_n\right\}$ is given by Proposition \ref{prop:expdelsingle}.
\end{thm}

Here again similar to the single hop case (Section \ref{sec:singlehop}) we see that finding a closed form expression for $P_s$ in terms of $D_1$ and $D_2$ is not possible due to the complicated expression for the joint probability of success on the two hops.  To gain more  insight into the dependence of $D_1$, and $D_2$ on $P_s$, and $C$, we derive a lower bound on $P_s$ as follows\footnote{Unlike the $N=1$ case, with $N >1$ we cannot obtain a simple and tight upper bound on the success probability. The difficulty in obtaining the upper bound is because the success event over the two hops is the complement of the union of the events \{failure on the first hop\}, and \{success on the first hop and failure on the second hop\}. Since the  \{success on the first hop\} is a decreasing event, and the \{failure on the second hop\} is an increasing event, FKG inequality cannot be used to upper bound the probability of success on two hops unlike the case for $N=1$.}. 
\subsection{Lower Bound On The Transmission Capacity}
Here we consider arbitrary number of hops $N$.
By definition  
 \[P_s  = P (\cap_{n=1,\ldots,N} \ \underbrace{\{\text{success in less than $D_n$ retransmissions on the $n^{th}$ hop}\}}_{S_{D_n}}).\] Event $S_{D_n}$ is a decreasing event, since for $\omega' \ge \omega$ ($\omega$ as defined in Example 1),  if $ {\mathbf 1}_{S_{D_n}}(\omega' ) =1 $ then automatically ${\mathbf 1}_{S_{D_n}}(\omega)=1$. Therefore, from the FKG inequality (Lemma \ref{lemfkg}), we get the following lower bound\footnote{The lower bound on the success probability corresponds to the case when the success event on each hop  are  independent. Since the spatial correlation coefficient of interference in a PPP is zero with path-loss model of 
 $d^{-\alpha}$ \cite{Ganti2009},  the derived lower bound is expected to be tight (also shown using simulations).}.
  
 \begin{lemma}\label{psuclbmultihop} ($D-R$ tradeoff of $N$ hop ad hoc network) The success probability in an ad hoc network with $N$ hop transmission  is lower bounded by $P_s \ge c^n \prod_{n=1}^N\left(1-(pq_{d_n}+1-p)^{D_n+1}\right)$. 
  \end{lemma}
\begin{proof}
$P_s = P\left(\cap_{n=1,\ldots,N}S_{D_n}\right)$.
Since $S_{D_n}$ is a decreasing event for each $n=1,\ldots, N$, $P_s \ge \prod_{n=1}^NP\left(S_{D_n}\right)$ from the FKG inequality. Result follows by substituting for $P\left(S_{D_n}\right)$ from  (\ref{eq:pucsingle}). 
 \end{proof}



The end-to-end transmissions/delay  is $M \bydef \sum_{n=1}^N M_n$, and by linearity of expectation $\bbE\{M\} = \sum_{n=1}^N\bbE\{M_n\}$. From (\ref{expdelay}), 
 \begin{equation}\label{delaymulti}
 \bbE\{M_n\} = \left[\frac{c\left(1-(pq_{d_n} + 1-p)^{D_n+1}\right)}{(1-q_{d_n})} +(D_n+1)(1-c) \right].
 \end{equation} 
 
 \begin{rem}\label{updelaymulti}
 Since $M_n \le D_n+1$, a simple upper bound on the expected end-to-end delay $\bbE\{M\}$ is $ \sum_{n=1}^N D_n +1 =  D+N$.  We will use this upper bound in next two sections to find the optimal $D_n$'s $(\sum_{n=1}^ND_n =D)$, and $N$ that maximize a lower bound on the transmission capacity. \end{rem}
 
  Using Lemma \ref{psuclbmultihop} and (\ref{delaymulti}), we obtain the following Theorem.
 \begin{thm}\label{thm:tcmulti} The transmission capacity of an ad hoc network with multi-hop transmission, and an end-to-end retransmission constraint of $D$ is lower bounded  by 
 \[C \ge \frac{\lambda R c^{n}\prod_{n=1}^N\left(1-(pq_{d_n}+1-p)^{D_n+1}\right)}{\sum_{n=1}^N\frac{c\left(1-(pq_{d_n} + 1-p)^{D_n+1}\right)}{(1-q_{d_n})} +(D_n+1)(1-c)}, \]bits/sec/Hz/$m^2$.
 \end{thm}

 \begin{rem}
 Note that an upper bound on the transmission capacity has been computed in  \cite{Andrews2009} for $D \rightarrow \infty$, and under the assumption that $SIR_t^n$ are independent $\forall \ t, n$, in which case $P_s=1$, and $\bbE\{M\} = \frac{N}{P_s^1}$.  Thus our result subsumes the result of \cite{Andrews2009}, since  with $SIR_t^n$ independent $\forall \ t, n$, $c=1$, and we have an equality in Lemma \ref{psuclbmultihop}, and Theorem \ref{thm:tcmulti}.
 \end{rem}

  {\it Discussion:} 
 In this section we first derived the D-R, and the T-D-R tradeoffs in an ad hoc network with multi-hop transmission from the source to its intended destination. The exact tradeoff expressions are quite complicated, and to get more insights we derived a lower bound on the success probability $P_s$, and the transmission capacity $C$. 
We showed that the end-to-end success probability is lower bounded by the product of the success probabilities on each hop.  Using the lower bound on $P_s$, we then derived a lower bound on the transmission capacity after exactly calculating the end-to-end delay to establish the T-D-R tradeoff.  
 Next, we derive  an analytically tractable lower bound on the transmission capacity using Remark \ref{updelaymulti}, and find the optimal $D_n$'s that maximize the lower bound. 


 \section{Optimal Per Hop Retransmissions}
 \label{sec:optD}
 In this section we derive a lower bound on the transmission capacity\footnote{The exact transmission capacity expression  is far too complicated for analysis.}, and then find the optimal  $D_n$'s that maximize the lower bound. From Remark \ref{updelaymulti}, 
 $\sum_{n=1}^N\bbE\{M_n\}  \le \sum_{n=1}^ND_n+1= D+N$, thus using the lower bound on $P_s$ (Lemma \ref{psuclbmultihop}), and the definition of transmission capacity (\ref{tcmulti})
 \begin{equation}\label{tclbsimple}C \ge \max_{D_n, \ \sum_{n=1}^ND_n \le D} \frac{\lambda R c^{n}\prod_{n=1}^N\left(1-(pq_{d_n}+1-p)^{D_n+1}\right)}{D+N}.\end{equation}

 \begin{prop}\label{prop:optD} The optimal $D_n^{\star}$'s that maximize the lower bound (\ref{tclbsimple}) on the transmission capacity satisfy $D^{\star}_n+1 = \frac{ln\left(\frac{\gamma}{ln({\hat q}_{d_n}) +\gamma}\right)}{ln({\hat q}_{d_n})}$,
 where $\gamma$ is such that $\sum_{n=1}^N D_n = D$.
 For equidistant hops $d_n = d/N, \ \forall \ n$, $D_{n}^{\star} = D/N$.
   \end{prop}
 \begin{proof} See appendix \ref{app:optD}.
\end{proof}
{\it Discussion:} 
In this section we first derived an analytically tractable lower bound on the transmission capacity, and then found sufficient conditions for finding the optimal $D_n$'s that maximize the derived lower bound. The  optimization function is  concave in $D_n$'s, and hence using the KKT conditions we derived the sufficient conditions for optimality.  For the special case of equidistant hops, $d_n= \frac{d}{N}$, we derived that equally distributing $D$ (the end-to-end delay constraint) among the $N$ hops,  maximizes the success probability. This result is quite intuitive in the sense that if for say hop $n$, $D_n < D_1=\ldots= D_{n-1}= D_{n+1}=  \ldots= D_N $, then the end-to-end success probability is dominated by the success probability of the $n^{th}$ hop, and is less than the success probability when $D_n=D_m, \ \forall \  n \ne m$.  
\section{Optimal Number of Hops $N$} 
\label{sec:optN} In this section we want to find the optimal number of equidistant hops $N$ that maximizes the lower bound (\ref{tclbsimple}) on the transmission capacity for a fixed $D$, with $D_n = \left\lfloor D/N\right\rfloor, \ \forall \ n$. Finding the optimal $N$ is a hard problem for arbitrary $\lambda$ and $D$. Next we show that in the sparse network regime $\lambda \rightarrow 0$, we can find an exact solution for the optimal $N$.


\begin{prop}\label{prop:optN} For a sparse network $\lambda\rightarrow 0$, $N=1$ maximizes\footnote{Note that throughout this paper we have assumed interference limited regime, and neglected the effects of additive  noise. The results of this section are unchanged  even while considering AWGN, since in that case $q_d = \left(1-\exp^{\left(-\frac{\beta\sigma^2d^{-\alpha}}{P}-\lambda c_1 \left(\frac{d}{N}\right)^2\beta^{\frac{2}{\alpha}}\right)}\right)$ \cite{Baccelli2006}, and once again for $\lambda \rightarrow 0$, we can show that transmission capacity lower bound is a decreasing function of $N$.}   the lower bound (\ref{tclbsimple}) transmission capacity for $p\approx 1$.
\end{prop}
\begin{proof} See appendix \ref{app:optN}.
\end{proof}

{\it Discussion:} In this section we showed that in a sparse network regime, it is optimal to transmit over a single hop. The physical interpretation of this result is that in a sparse network with few interferers, the decrease in transmission capacity due to the end-to-end delay (linear in $N$) outweighs the increase in transmission capacity due to the reduced per hop distance $\left(\frac{d}{N}\right)$.  Our result is in agreement with  \cite{Andrews2009}, where the transmission capacity (eq. 12) is a decreasing function of $N$ for small values of $\lambda$.  

\section{Simulations}
\label{sec:sim}
In all the simulation results we use  $\alpha =3$,  $\beta =3$ corresponding to 
$R=2$ bits/sec/Hz, $p=1/2$, and  $\lambda = 0.1$ (expect Fig. \ref{fig:tchophigh}). In Figs. \ref{fig:dr}, and \ref{fig:dr2hop}, we plot the success probability $P_s$ as a function of the maximum number of retransmissions $D$ for single hop, and two-hop communication, respectively. 
We also plot the derived upper and lower bounds on the success probability. We can see that the upper and lower bound are tight. In Figs. \ref{fig:tcdelaysymm},  and \ref{fig:tcdelayassym}, we plot the transmission capacity, and the derived lower bound for two-hop communication $N=2$, with respect to  $D_1$, with  $D=4$, for equidistant hops $d_1=d_2 =1m$, and non equidistant hops $d_1=0.5m, \  d_2=1.5 m$, respectively.  The transmission capacity (simulated and the lower bound) is maximized at $D_1=D_2=2$ for $d_1=d_2 =1m$, and $D_1=1, \ D_2 = 3$ for $d_1=0.5m, \  d_2=1.5 m$ which is in accordance with Proposition \ref{prop:optD}. In Figs. \ref{fig:tchoplow}, and \ref{fig:tchophigh}, we plot the transmission capacity as the function of the number of hops $N$ with $D=10$ for transmission density $\lambda = 0.1$, and $\lambda =0.5$, respectively. For small $\lambda =0.1$ as derived in Proposition \ref{prop:optN} optimal $N=1$, however, as we increase to $\lambda=0.5$, that is no longer true (also shown in \cite{Andrews2009}), and the transmission capacity is not monotonic in $N$.
\appendices
\section{}\vspace{-0.4in}
\label{app:jointprob}
\begin{eqnarray}
\nonumber 
 P\left(
SIR_{m}<\beta_{,m=1,2,\ldots,k,},
 \  SIR_j \ge \beta  \right) & = & P\left(\frac{d^{-\alpha}|h_{00}^1|^2}{I_{\Phi_1}}< \beta, \ldots, \frac{d^{-\alpha}|h_{00}^{k}|^2}{I_{\Phi_{k+1}}}< \beta, \ \frac{d^{-\alpha}|h_{00}^{j}|^2}{I_{\Phi_{j}}}\ge \beta\right), \\\nonumber
  &\stackrel{(a)}=& \bbE_{\Phi_{\ell}, \Phi_{j}, h_{0n}^{\ell}} 
  \left\{\prod_{\ell=1}^{k} \left(1- \exp\left(-\frac{\beta I_{\Phi_{\ell}}}{d^{-\alpha}}\right)\right)\left(\exp\left(-\frac{\beta I_{\Phi_{j}}}{d^{-\alpha}}\right)\right)\right\},\\\nonumber
   &\stackrel{(b)}=& 
 \bbE_{\Phi_{\ell},  \Phi_{j},h_{0n}^{\ell}} \left\{\prod_{\ell=1}^{k}
 \left(
 1- \exp
 \left(
 -\frac{
 \beta \sum_{T_n^{\ell} \in \Phi_{\ell} \backslash \{T_0\}} \mathbf{1}_{T_n^{\ell}} d_{Tn}^{-\alpha}|h^{\ell}_{0n}|^2
 }
 {
 d^{-\alpha}
 }
 \right)
 \right) 
 \right. \\\nonumber
 && 
  \left.
  \exp
 \left(
 -\frac{
 \beta \sum_{T_n^{j} \in \Phi_{j} \backslash \{T_0\}} \mathbf{1}_{T_n^j} d_{Tn}^{-\alpha}|h^{j}_{0n}|^2
 }
 {
 d^{-\alpha}
 }
 \right)\right\},
 \end{eqnarray}
  \begin{eqnarray}\nonumber
   &\stackrel{(c)}=& 
   \bbE_{\Phi} 
   \left\{ \prod_{\ell=1}^{k} 
  \left(1- \prod_{x \in \Phi \backslash \{T_0\}}\left( \frac{p}{1+ d^{\alpha}\beta x^{-\alpha}} + 1-p \right)   \right)  \prod_{x \in \Phi\backslash \{T_0\}}\left( \frac{p}{1+ d^{\alpha}\beta x^{-\alpha}} + 1-p \right) \right\}, \\\nonumber
  &\stackrel{(d)}=& 
   \bbE_{\Phi} \left\{X_{\Phi}(1-X_{\Phi})^{k}\right\}, \\\nonumber
   &=& \sum_{\ell=0}^{k} (-1)^{\ell}{k \choose \ell }\bbE_{\Phi} \left\{X_{\Phi}^{\ell+1}\right\}, \\\nonumber
   &\stackrel{(e)}=& \sum_{\ell=0}^{k} (-1)^{\ell}{k\choose \ell }  \exp\left( -\lambda \int_{\bbR^2} 1- \left(\frac{p}{1+ d^{\alpha}\beta x^{-\alpha}} + 1-p \right)^{\ell+1} {\mathbf d}x \right),    \end{eqnarray}
 where $(a)$ follows by  taking the expectation with respect to $h_{00}^t, \ t=1,\ldots, k+1, \ j$,  since $|h_{00}^t|^2$ are independent and exponentially distributed, $(b)$ follows from definition of $I_{\Phi_{\ell}}$, $(c)$
 follows by  taking the expectation with respect to $h_{0n}^{\ell}$ and ALOHA, $(d)$ follows by defining  $X_{\Phi} = \prod_{x \in \Phi \backslash \{T_0\}}\left( \frac{p}{1+ d^{\alpha}\beta x^{-\alpha}} + 1-p \right)$, and $(e)$ follows from the probability generating function of PPP \cite{Stoyan1995}.   
\vspace{-0.3in}
\section{}
\label{app:upboundPs}
Recall that each transmitter retransmits with probability $p$ in each time slot. Let ${\cal S}_0$ make $k$ attempts to transmit the packet to ${\cal D}_0$, $k=1,2,\ldots, D+1$. Then the event 
$\{$success in at most $D$ retransmissions$\}$ is also equal to the complement of the event $\{\text{failures in all $k$ attempts}\}$ for $k=1,2,\ldots, D+1$.  Thus, 
\begin{eqnarray}\nonumber
P_s &=& 1 - \sum_{k=1}^{D+1} {D +1 \choose k} p^k(1-p)^{D+1-k} P(\text{failure in $k$ attempts}),\\ \label{uppsuc1}
        &=& 1-  \sum_{k=1}^{D+1} {D +1 \choose k} p^k(1-p)^{D+1-k} P\left(
SIR_1 <  \beta, 
\ \ldots,  \ SIR_{k} <  \beta  \right),
 \end{eqnarray}
 since each $SIR_t$ is identically distributed, it does not matter where those $k$ failures happen.
 Similar to example \ref{exm:incevent}, it easily follows that $\{SIR_1 < \beta\}$ is an increasing event. Thus, using the FKG inequality, 
  \begin{eqnarray}\nonumber
P\left(SIR_1 <  \beta, 
\ \ldots,  \ SIR_{k} <  \beta  \right)& \ge& P\left(
SIR_1 <  \beta\right) \ldots P\left(SIR_{k} <  \beta  \right),\\ \label{plb}
& = &P\left(SIR_1 <  \beta  \right)^k,
 \end{eqnarray}
 since $P\left(SIR_t <  \beta  \right)= P\left(SIR_m <  \beta  \right)$, $t\ne m$.
 Let $q= P\left(SIR_1<  \beta  \right)$. 
 From \cite{Baccelli2006}, 
 \begin{equation}\label{pout}
 q= P\left(SIR_1<  \beta  \right)= 1-\exp \left(-\frac{\lambda 2 \pi^2 d^2 \beta^{
 \frac{2}{\alpha}}Csc\left(\frac{2\pi}{\alpha}\right)}{\alpha}\right) \bydef  1-\exp \left(-\frac{\lambda c_1 d^2 \beta^{
 \frac{2}{\alpha}}}{\alpha}\right).\end{equation}
Substituting (\ref{plb}) into (\ref{uppsuc1}), 
$P_s \le  1 - \sum_{k=1}^{D+1} {D +1 \choose k} p^k(1-p)^{D+1-k} q^k = 1- (pq+1-p)^{D+1}$.

\section{}
\label{app:lboundPs}
From (\ref{psucdef1})
$P_{s} 
= \sum_{j=1}^{D+1} \sum_{k=0}^{j-1}  \ {j-1 \choose k} p^k(1-p)^{j-1-k} p
P\left(SIR_1 <  \beta, 
\ \ldots,  \ SIR_{k} <  \beta,
 \  SIR_j \ge \beta  \right)$.

Note that 
\begin{eqnarray*}
P\left(SIR_1 <  \beta, 
\ \ldots,  \ SIR_{k} <  \beta,
 \  SIR_j \ge \beta  \right) &=& 
P\left(SIR_1 <  \beta, 
\ \ldots,  \ SIR_{k} <  \beta  \right)\\ 
&& P\left(SIR_j \ge \beta \ | \ SIR_1 <  \beta, 
\ \ldots,  \ SIR_{k} <  \beta \right) , \\ 
&\ge & q^{k} P\left(SIR_j \ge \beta \ | \ SIR_1 <  \beta, 
\ \ldots,  \ SIR_{k} <  \beta \right), \end{eqnarray*}
since from the FKG inequality $P\left(SIR_1 <  \beta, 
\ \ldots,  \ SIR_{k} <  \beta  \right) \ge q^{k}$. 
Hence 
\begin{eqnarray*}
P_{s} 
&\ge & \sum_{j=1}^{D+1} \sum_{k=0}^{j-1}  \ {j-1 \choose k} p^k(1-p)^{j-1-k} p q^k P\left(SIR_j \ge \beta \ | \ SIR_1 <  \beta, 
\ \ldots,  \ SIR_{D} <  \beta \right)
\end{eqnarray*}
From \cite{Ganti2009} 
\[P\left(SIR_j \ge \beta \ | \ SIR_1 <  \beta, 
\ \ldots,  \ SIR_{k} <  \beta \right) \ge P\left(SIR_j \ge \beta \ | \ SIR_1 <  \beta, 
\ \ldots,  \ SIR_{k+1} <  \beta \right) \] for any $k+1<j$. 
Therefore, since $SIR_j$ are identically distributed for all $j$, 
\[P\left(SIR_j \ge \beta \ | \ SIR_1 <  \beta, 
\ \ldots,  \ SIR_{k} <  \beta \right) \ge P\left(SIR_{D+1} \ge \beta \ | \ SIR_1 <  \beta, 
\ \ldots,  \ SIR_{D} <  \beta \right) \] 
Thus, we get the following lower bound on $P_s$ 
\begin{eqnarray}\nonumber
P_{s}&\le & P\left(SIR_{D+1} \ge \beta \ | \ SIR_1 <  \beta, 
\ \ldots,  \ SIR_{D} <  \beta \right)\sum_{j=1}^{D+1} \sum_{k=0}^{j-1}  \ {j-1 \choose k} p^k(1-p)^{j-1-k} p q^k , \\\nonumber
&= & P\left(SIR_{D+1} \ge \beta \ | \ SIR_1 <  \beta, 
\ \ldots,  \ SIR_{D} <  \beta \right) \frac{1- (pq+1-p)^{D+1}}{1-q}. \label{lbpsuc}
  \end{eqnarray}

  \section{}
\label{app:comparison}
By definition
\begin{small}
 \begin{eqnarray*}
\nonumber 
 P\left(
SIR_{D+1} \ge  \beta \ | \
SIR_ 1< \beta \ \ldots,  \ SIR_{D} <  \beta\right)  & =& \frac{P\left(
SIR_1 <  \beta, 
\ \ldots,  \ SIR_{D} <  \beta,
 \  SIR_{D+1} \ge \beta  \right)}{P\left(
SIR_1 <  \beta, 
\ \ldots,  \ SIR_{D} <  \beta, \right)}, \\
&\stackrel{(a)}{=}& 
\frac{\sum_{\ell=0}^{D} (-1)^{\ell}{D\choose \ell }\exp\left( -\lambda \int_{\bbR^2} 1- \left(\frac{p}{1+ d^{\alpha}\beta x^{-\alpha}} + 1-p \right)^{\ell+1} {\mathbf d}x \right)}{\sum_{\ell=0}^{D} (-1)^{\ell}{D\choose \ell }\exp\left( -\lambda \int_{\bbR^2} 1- \left(\frac{p}{1+ d^{\alpha}\beta x^{-\alpha}} + 1-p \right)^{\ell} {\mathbf d}x \right)}, 
 \end{eqnarray*} 
\end{small}
where $(a)$ follows from Appendix \ref{app:jointprob}.
Hence for $D=2$, by computing the integral for $\ell=1,2$, with $c_2\bydef p \lambda 2 \pi d^2 \beta^{
 \frac{2}{\alpha}}$
\begin{small}
\begin{eqnarray*}
\nonumber 
 P\left(
SIR_{3} \ge  \beta \ | \
SIR_ 1< \beta, \ SIR_{2} <  \beta\right)  & =& \frac{\exp \left(-\frac{\pi c_2 Csc\left(\frac{2\pi}{\alpha}\right)}{\alpha}\right) - 2 
\exp\left(
-c_2 \left(2 - p\pi \frac{\alpha-2}{\alpha^2}\pi Csc\left(\frac{2\pi}{\alpha}\right)\right)\right)}
{1-2\exp \left(-\frac{c_2 \pi Csc\left(\frac{2\pi}{\alpha}\right)}{\alpha}\right) + \exp\left(
-c_2\left(2 - p\pi \frac{\alpha-2}{\alpha^2}\pi Csc\left(\frac{2\pi}{\alpha}\right)\right)\right)} \\
&&+ \frac{\exp\left(-\frac{c_2 \left(-3 \alpha(-2+p)p+ 2p^2+\alpha^2(3+p(p-3))\right)Csc\left(\frac{2\pi}{\alpha}\right)}{\alpha^3}\right)}
{1-2\exp \left(-\frac{c_2 \pi Csc\left(\frac{2\pi}{\alpha}\right)}{\alpha}\right) + \exp\left(
-c_2 \left(2 - p\pi \frac{\alpha-2}{\alpha^2}\pi Csc\left(\frac{2\pi}{\alpha}\right)\right)\right)},\\
&\approx&  \exp(-\lambda2\pi p \beta^{\frac{2}{\alpha}} c_1d^2)= 1-q. \end{eqnarray*}
\end{small}
Similar conclusion can be drawn for $D=3$ using $(a)$ and by substituting 
\begin{small}
\begin{eqnarray*}
\exp\left( -\lambda \int_{\bbR^2} 1- \left(\frac{p}{1+ d^{\alpha}\beta x^{-\alpha}} + 1-p \right)^{4} {\mathbf d}x \right) &=& 
\exp\left(-\frac{
c_2 \pi \left(
-12 \alpha(-2+p)p^2 \right)Csc\left(\frac{2\pi}{\alpha}\right)}{3\alpha^4}\right. \\ 
&& - \left. \frac{c_2 \pi Csc\left(\frac{2\pi}{\alpha}\right)\left(4p^3-3\alpha^3(p-2)(2+p(p-2)) + 
\right)}{3\alpha^4}\right. \\
&& \left.- \frac{c_2 \pi Csc\left(\frac{2\pi}{\alpha}\right)\left( \alpha^2p(36 + p(11p-36))
\right)}{3\alpha^4}\right).
\end{eqnarray*}
\end{small}

\vspace{-0.5in}

\section{}
\label{app:optD}
Let ${\hat q}_{d_n} \bydef 1-p+pq_{d_n}$, from (\ref{tclbsimple}) the objective function is
 \[ \max_{D_n, \sum_{n=1}^ND_n=D } \prod_{n=1}^N (1-({\hat q}_{d_n})^{D_n+1}).\]
 Since $\log$  is  a monotone function, an equivalent problem is 
 $\max_{D_n, \sum_{n=1}^ND_n=D }  \sum_{n=1}^N\ln(1-{\hat q}_{d_n}^{D_n+1})$. It is easy to verify that the objective function is a concave function in $D_n$.
 Using Lagrange multiplier $\gamma$, we can write the Lagrangian as 
 \[{\cal L} =   \sum_{n=1}^N\ln(1-{\hat q}_{d_n}^{D_n+1}) + \gamma(\sum_{n=1}^ND_n- D).\]
 Differentiating with respect to $D_n$, and equating it to zero, we have 
 \[\frac{d{\cal L}}{dD_n} = \frac{- \ln ({\hat q}_{d_n}){\hat q}_{d_n}^{D_n+1}}{1-{\hat q}_{d_n}^{D_n+1}}+\gamma = 0, \implies \ D_n+1 = \frac{\ln\left(\frac{\gamma}{\ln({\hat q}_{d_n}) +\gamma}\right)}{\ln({\hat q}_{d_n})}\]
Finding an explicit solution for the optimal $\gamma$ is analytically intractable, hence we need to use an iterative algorithm to find optimal $\gamma$,  at each step $\gamma$ is increased if $\sum_{n=1}^ND_n< D$, or decreased if $\sum_{n=1}^ND_n> D$, similar to the Waterfilling solution \cite{Cover2004}. For equidistant hops   $d_n=d/N$, ${\hat q}_d\bydef {\hat q}_{d_n} \ \forall \  n$, the optimal 
$\gamma =
\frac{ \ln({\hat q}_d)
\exp{
\left(
\frac{ \ln({\hat q}_d)(D+N) }{N} 
\right)
}
}
{
1-\exp{
\left(
\frac{ \ln({\hat q}_d)(D+N) }{N} 
\right)
}
}
 $, and $D^{\star}_n= \frac{D}{N}, \forall \ n$ if $D$ is a multiple of $N$. 

\section{}
\label{app:optN}
Using $ d_n = d/N, \ \forall \ n, \implies q_{d}=q_{d_n}$, and $D_n = \left\lfloor D/N\right\rfloor, \forall  \ n$, for $p\approx 1$,  the lower bound on the transmission capacity  is  
$C \ge  \frac{\lambda R \prod_{n=1}^N \left(1-q_d^{\left\lfloor D/N\right\rfloor+1}\right)}{D+N}$.
Recall from (\ref{pout}) that  $q_d =  (1-\exp^{-\lambda c_1 \left(\frac{d}{N}\right)^2\beta^{\frac{2}{\alpha}}})$. Thus the optimization function is 
$\max_{N}  \frac{\left(1-(1-\exp^{-c_1\lambda\left(\frac{d}{N}\right)^2\beta^{\frac{2}{\alpha}}})^{\left\lfloor D/N\right\rfloor+1}\right)^N}{D+N}$.
Using the Taylor series expansion of $\exp^{\left(-\lambda\left(\frac{d}{N}\right)^2\beta^{\frac{2}{\alpha}} \right)}$ for $\lambda \rightarrow 0$, and keeping only the first two terms,  the objective function is
\begin{eqnarray*}
&=&  \max_{N} \frac{\left(1-\left(c_1\lambda\left(\frac{d}{N}\right)^2\beta^{\frac{2}{\alpha}}\right)^{\left\lfloor D/N\right\rfloor+1}\right)^N}{D+N}, \\ 
&=& \max_{N} \frac{1-N\left(c_1\lambda d^2 \beta^{\frac{2}{\alpha}}\left(\frac{1}{N^2}\right)\right)^{\left\lfloor D/N\right\rfloor+1}}{D+N} + {\cal O}(\lambda^2), \\
&=& \max_{N}\frac{1}{D+N}-\frac{\left(c_1 \lambda d^2 \beta^{\frac{2}{\alpha}}\right)^{\left\lfloor D/N\right\rfloor+1} \left(\frac{1}{N}\right)^{\frac{D}{N}}}{D+N}+ {\cal O}(\lambda^2).
\end{eqnarray*}
Note that for small $\lambda d^2 \beta^{\frac{2}{\alpha}}$ for which the Taylor series expansion is valid, this expression is a  a decreasing 
function of $N$, thus, $N=1$ maximizes the success probability for  a sparse network $\lambda \rightarrow 0$.

\bibliographystyle{IEEEtran}
\bibliography{IEEEabrv,Research}
\newpage
\begin{figure}
\centering
\includegraphics[width=3in]{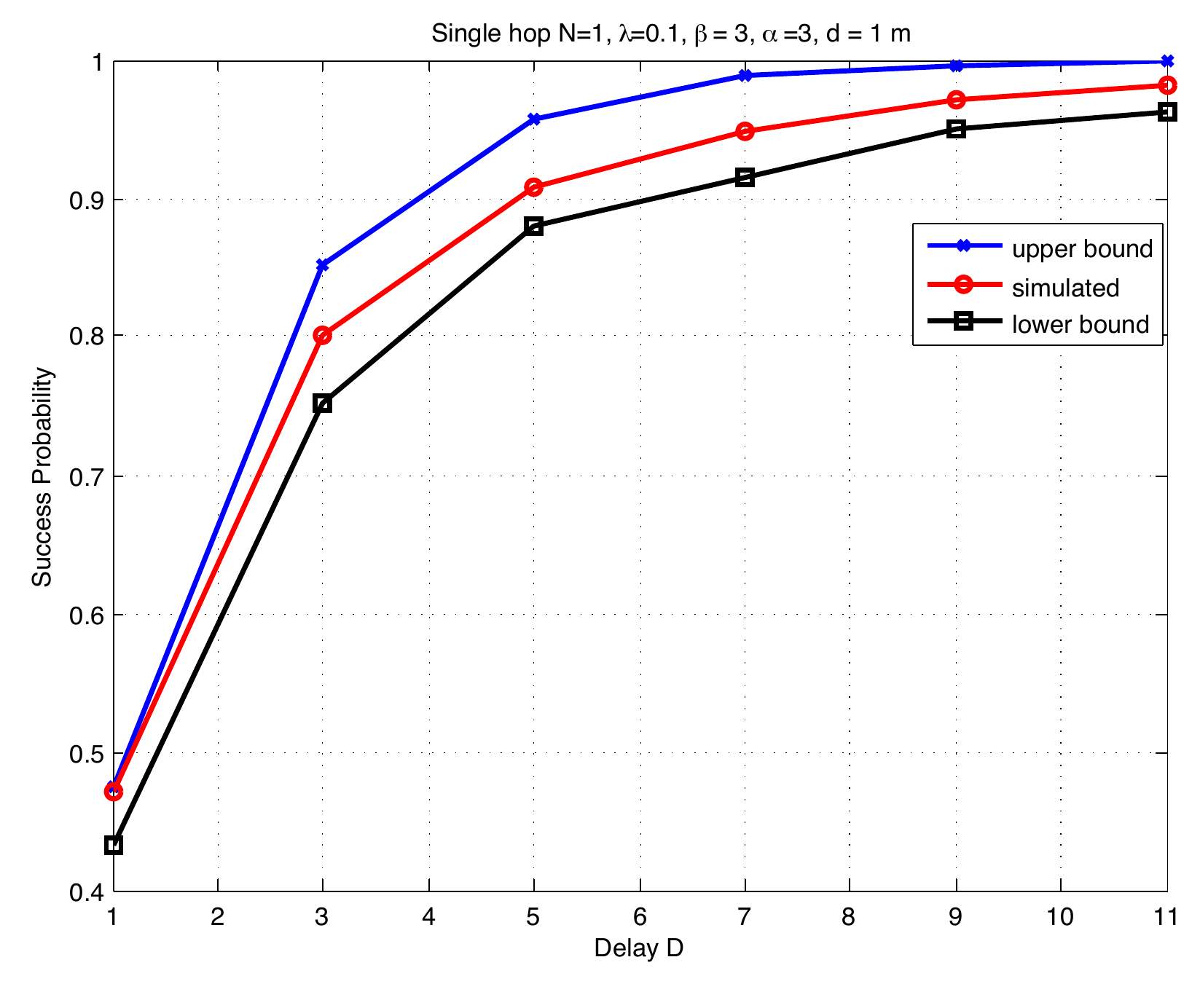}
\caption{Success probability as a function of D for N=1.}
\label{fig:dr}
\end{figure}

\begin{figure}
\centering
\includegraphics[width=3.5in]{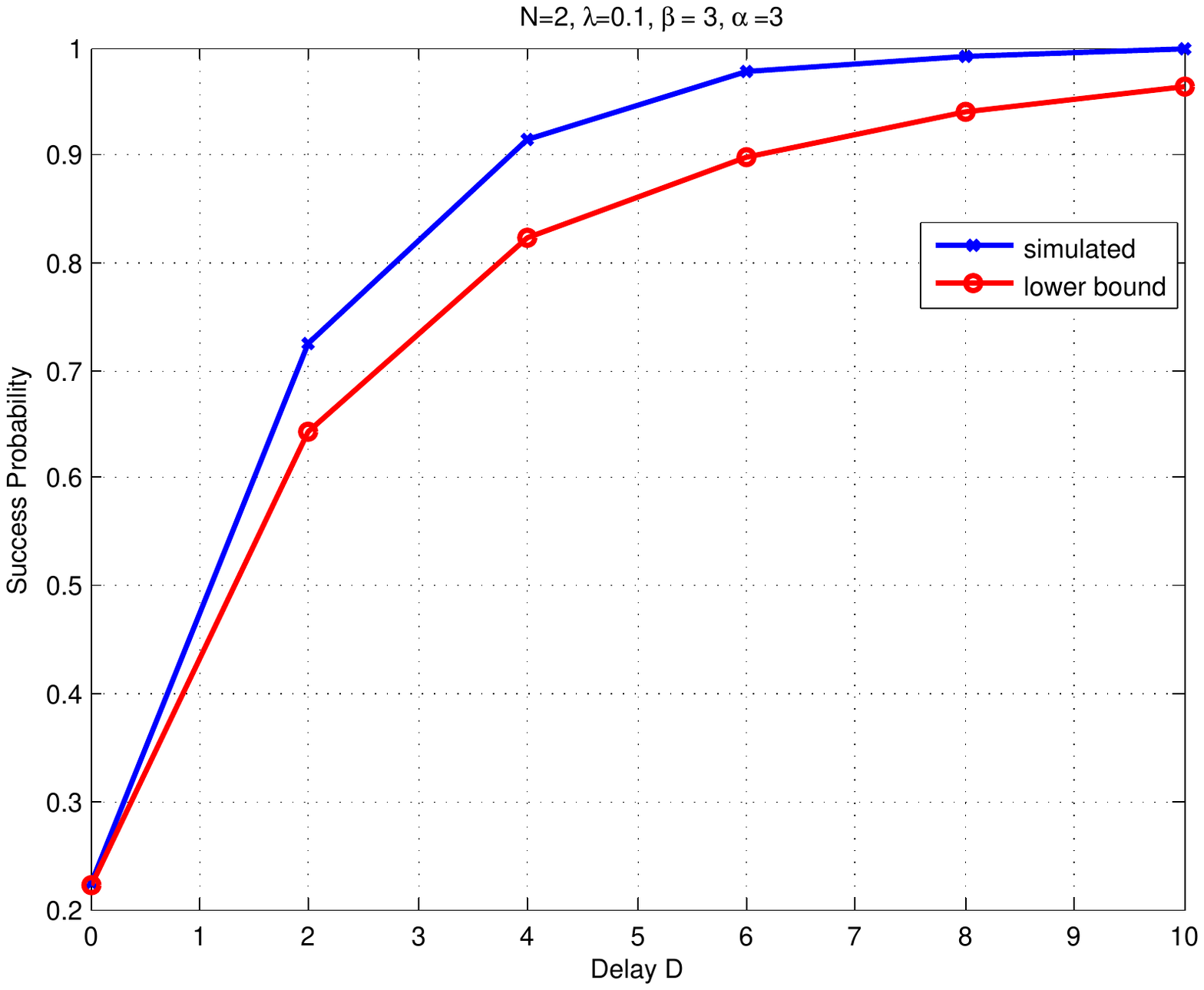}
\caption{Success probability as a function of D for N=2.}
\label{fig:dr2hop}
\end{figure}

\begin{figure}
\centering
\includegraphics[width=3in]{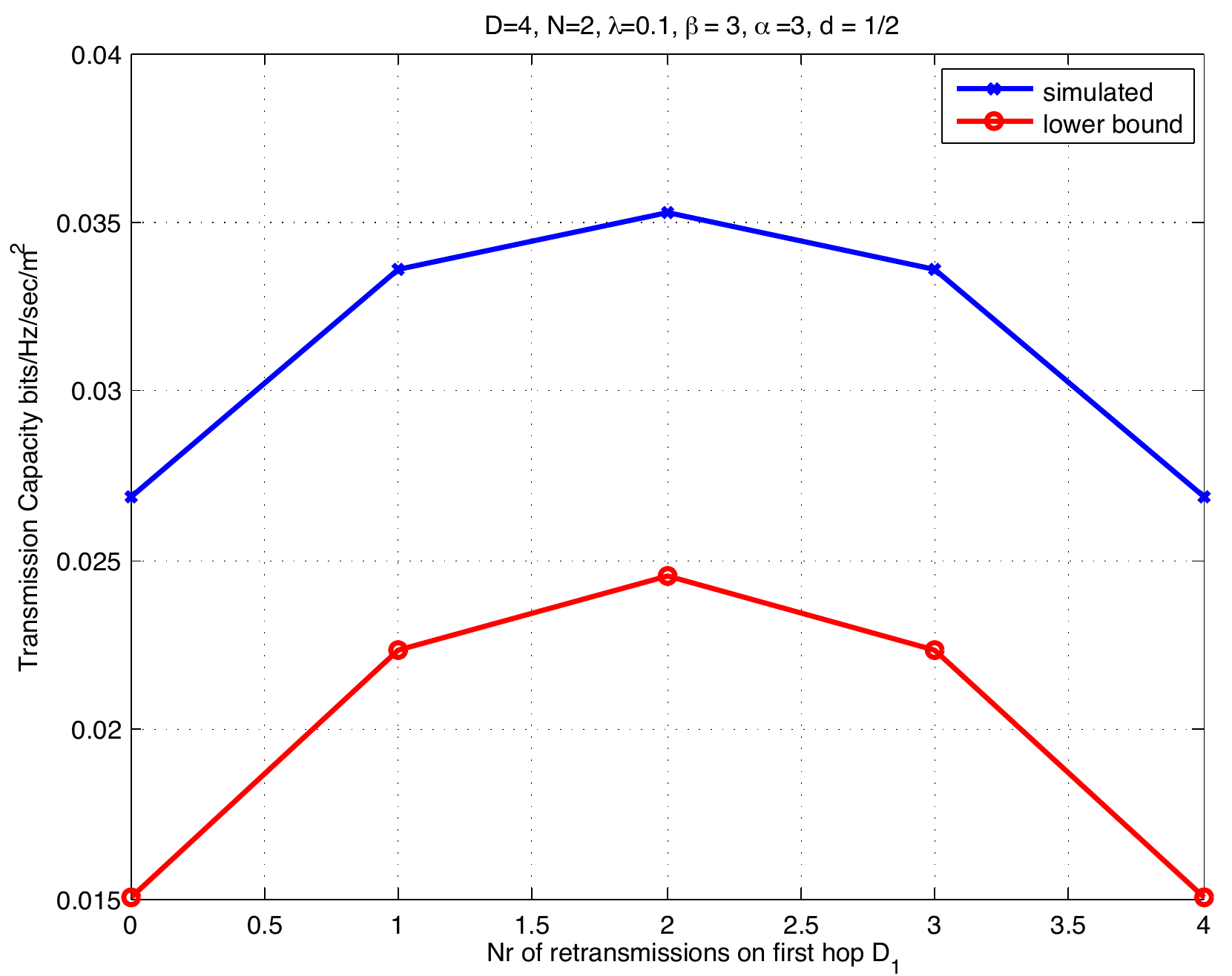}
\caption{Transmission capacity as a function of $D_1$ with $D=4$ for equidistant hops.}
\label{fig:tcdelaysymm}
\end{figure}

\begin{figure}
\centering
\includegraphics[width=3in]{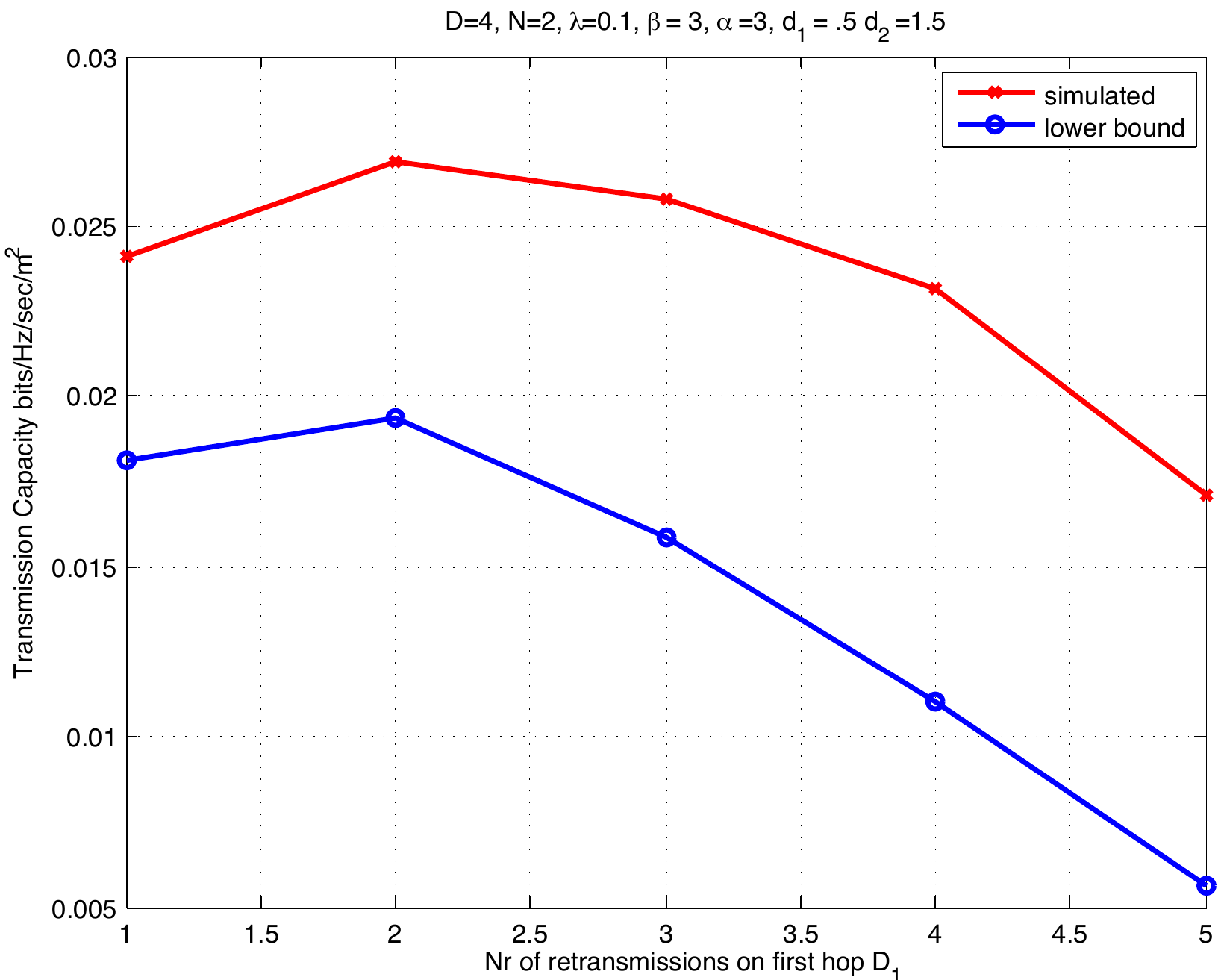}
\caption{Transmission capacity as a function of $D_1$ with $D=4$ for non equidistant hops.}
\label{fig:tcdelayassym}
\end{figure}

\begin{figure}
\centering
\includegraphics[width=3in]{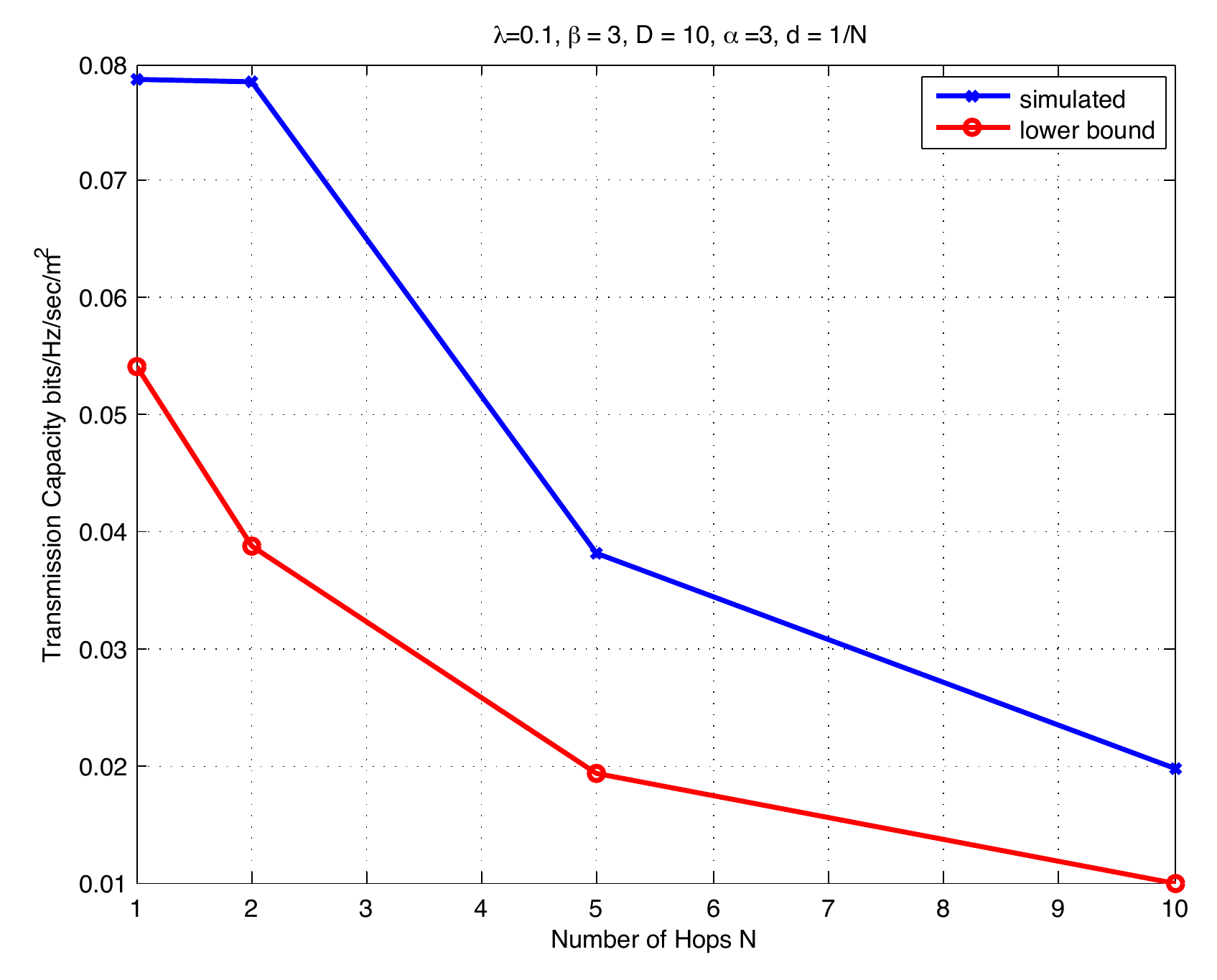}
\caption{Transmission capacity as a function of number of hops $N$ for $\lambda =0.1$.}
\label{fig:tchoplow}
\end{figure}

\begin{figure}
\centering
\includegraphics[width=3in]{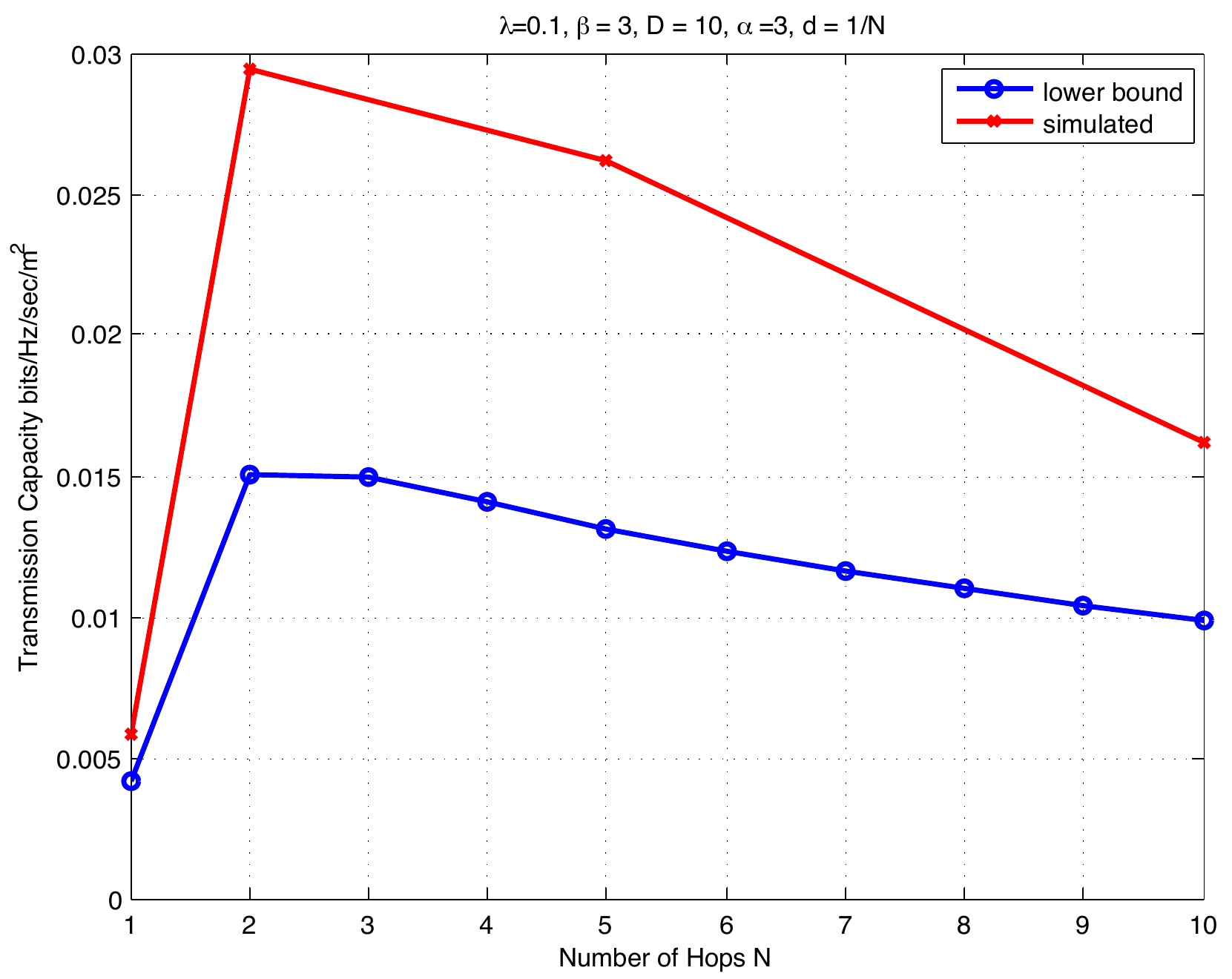}
\caption{Transmission capacity as a function of number of hops $N$ for $\lambda =0.5$.}
\label{fig:tchophigh}
\end{figure}

\end{document}